\documentclass[showpacs,aps,superscriptaddress,twocolumn]{revtex4}

\def\ie{{\it i.e.\ }}

\def\f0f2{f$_{0}$F$_{2}$}

\def\ch2{$\chi^2$}

\def\msol{M$_\odot$ }

\def\f0f2{f$_{0}$F$_{2}$}

\def\h0{$H_0$}
\def\q0{$q_0$}

\def\CO{$\Omega_{\rm CO}$ }

\newcommand{\lta}{\mbox{\small\raisebox{-0.6ex}{$\,\stackrel
{\raisebox{-.2ex}{$\textstyle <$}}{\sim}\,$}}}
\newcommand{\gta}{\mbox{\small\raisebox{-0.6ex}{$\,\stackrel
{\raisebox{-.2ex}{$\textstyle >$}}{\sim}\,$}}}   

\begin{document}

\title{Limits on the cosmological abundance of supermassive compact
objects from a search for multiple imaging in compact radio sources}

\author{P.N. Wilkinson}
\affiliation{University of Manchester, Jodrell Bank Observatory,
Macclesfield, Cheshire SK11 9DL, UK}
\author{D.R. Henstock}
\affiliation{University of Manchester, Jodrell Bank Observatory,
Macclesfield, Cheshire SK11 9DL, UK}
\author{I.W.A. Browne}
\affiliation{University of Manchester, Jodrell Bank Observatory,
Macclesfield, Cheshire SK11 9DL, UK}
\author{A.G. Polatidis} 
\affiliation{University of Manchester, Jodrell Bank Observatory,
Macclesfield, Cheshire SK11 9DL, UK}
\affiliation{Onsala Space Observatory, Chalmers Institute of
Technology, S-43992 Onsala, Sweden}
\author{P. Augusto}
\affiliation{University of Manchester, Jodrell Bank Observatory,
Macclesfield, Cheshire SK11 9DL, UK}
\affiliation{Universidade da Madeira, Dep.\ Matem\'atica, Caminho da
Penteada, 9050 Funchal,Portugal}
\author{A.C.S. Readhead} 
\affiliation{California Institute of Technology, Pasadena, California, 91125}
\author{T.J. Pearson}
\affiliation{California Institute of Technology, Pasadena, California, 91125}
\author{W. Xu}
\affiliation{California Institute of Technology, Pasadena, California, 91125}
\author{G.B. Taylor} 
\affiliation{California Institute of Technology, Pasadena, California, 91125}
\affiliation{National Radio Astronomy Observatory, Socorro New Mexico, 87801}
\author{R.C. Vermeulen}
\affiliation{California Institute of Technology, Pasadena, California, 91125}
\affiliation{Netherlands Foundation for Radioastronomy, P.O. Box 2,
7990 AA Dwingeloo, The Netherlands Dwingeloo, The Netherlands}

\date{\today}

\begin{abstract}
Using Very Long Baseline Interferometry we have searched a sample of
300 compact radio sources for examples of multiple imaging produced by
gravitational lensing; no multiple images were found with separations
in the angular range 1.5--50 milliarcsec. This null result allows us
to place a limit on the cosmological abundance of intergalactic
supermassive compact objects in the mass range $\sim 10^{6}$ to $\sim
10^{8}$M$_{\odot}$; such objects cannot make up more than $\sim 1\%$
of the closure density (95\% confidence). A uniformly distributed
population of supermassive black holes forming soon after the Big Bang
do not, therefore, contribute significantly to the dark matter content
of the Universe. 
\end{abstract}


\maketitle

\section{Introduction}

The possibility that a first generation of objects with masses
comparable with those of globular clusters formed prior to galaxies
has long been recognised \cite{Pee68,Car84}. Such Jeans--mass ($\sim
10^{6.5}$ \msol) objects forming shortly after the decoupling of
matter and radiation in the early universe could have evolved to black
holes and it is possible that some of the dark matter could be in
this, difficult to detect, form \cite{Car94,Car99}. Building on these
ideas Gnedin \& Ostriker \cite{Gne92} and Gnedin, Ostriker \& Rees
\cite{Gne95} explored a cosmogonic model in which the baryonic density
($\Omega_{b} \sim 0.15$) is up to an order of magnitude higher than
that inferred from primordial nucleosynthesis with the ``excess''
baryons being lost in the collapse of such Jeans--mass objects. It has
also been conjectured that relic massive black holes might provide the
seeds for quasars \cite{Fuk96}.

Press \& Gunn \cite{Pre73} developed the idea of detecting
supermassive compact objects (CO) by their gravitational lensing
effects well before the actual discovery of gravitational lenses in
1979. They showed that in a universe filled with a mass density \CO
$\sim 1$, the probability of a distant source being detectably lensed
by a supermassive CO is of order unity, while for \CO $< 1$, the
probability decreases in direct proportion to the mass density.  From
this they drew the important conclusion that the fraction of distant
galaxies that is lensed by CO directly measures \CO and, ignoring
angular resolution effects, is independent of the mass $M_{CO}$ of the
lenses. The latter property is simply understood. A given value of \CO
can be made up of a large number of low--mass objects or a small
number of high--mass ones; hence the number density $n$ of CO of a
particular mass is proportional to $1/M_{CO}$. For point masses the
gravitational lensing cross--section $\sigma \propto M_{CO}$
\cite{Tur84} and hence the path length to lensing $(1/n\sigma)$ is
independent of the lens mass. However, the average image separation
measures $M_{CO}$ directly and is approximately independent of
$\Omega_{CO}$. These ideas were further developed by Nemiroff
\cite{Nem89} and by Nemiroff \& Bistolas \cite{Nem90}.

Since for a lens at a cosmologically significant distance the image
separation is $ \sim 2 \times 10^{-6} (M_{CO}/M_{\odot})^{1/2}$
arcseconds, searching for Jeans--mass CO requires the milliarcsec
(mas) resolution of Very Long Baseline Interferometry
(VLBI). Kassiola, Kovner \& Blandford (\cite{Kas91}; hereafter KKB),
used the lack of ``millilensed'' images in published VLBI maps of 48
compact radio sources, to place an upper limit \CO $< 0.4$ (99.7\%
confidence) in the mass range $10^7$--$10^9$ M$_{\odot}$. In a
closely--related search Nemiroff \cite{Nem93} used the lack of
secondary ``echoes'' of gamma ray bursts, which would also arise from
millilensing, to rule out a closure density of supermassive CO. In
this paper we describe a millilens search based on high-quality VLBI
maps of 300 sources which extends KKB's lower mass range to $\sim
10^6$ \msol and enables us to push their limit on \CO for Jeans--mass
objects down by a factor 15. This is the most stringent limit to date
on \CO for uniformly--distributed Jeans--mass objects and we discuss
some cosmological implications of this result.

\section{Selection of Lens Candidates}

The parent sample was drawn from catalogues of compact radio sources
which have been systematically observed by the authors with
intercontinental VLBI arrays at a frequency of 5 GHz and a resolution
$\sim 1$ mas; a significant fraction of these sources was also
observed at 1.6 GHz. The catalogues, radio images and full description
of the observations can be found in the following papers: the Pearson
\& Readhead VLBI Survey (PR; \cite{Pea88}); the first Caltech--Jodrell
Bank VLBI Survey (CJ1; \cite{Pol95,Tha95,Xu95}); the second
Caltech--Jodrell Bank VLBI Survey (CJ2; \cite{Tay94,Hen95}). From
these samples a large sub--set of 300 sources was selected on the
basis of a ``flat'' integrated radio spectrum between 1.4 GHz and 5
GHz.  In this context ``flat'' implies a source whose spectral index
$\alpha \geq -0.5$, where the flux density at a frequency $\nu$ is
proportional to $\nu^{\alpha}$. The sample thus selected contains
nearly all the strongest flat--spectrum radio sources in the sky above
declination $35^{\circ}$.

Flat-spectrum radio sources are best--suited to a millilens search
because they tend to be dominated by a single compact (sub--mas)
``core'', which originates close to the massive central black hole in
the nucleus of the background galaxy.  The presence of {\it two}
compact components (putative images) then provides a simple diagnostic
of gravitational lensing by a point--like mass and genuine images will
have well understood, simply--related, properties \cite{Tur84}. However
great care has to be taken in choosing lens candidates since compact
radio sources have a range of intrinsic structures which can confuse
the selection process. Guided by a series of simulations we decided
only to look for cases of multiple imaging of the compact core. It is
much harder to be certain of recognising lensing effects on
lower--brightness emission regions.  We adopted the following
quantitative selection criteria based on the Gaussian models listed in
the original VLBI survey papers:

{\it A candidate has to have two or more compact ($<$ 1 mas)
components and the secondary component should be smaller than the
primary.} Because of the surface--brightness conserving property of
lensing, weaker images are smaller images. However, conservative
allowance was made for Gaussian model-fitting uncertainties by
accepting some candidates in which the secondary had an area up to
twice, and in a few cases three times, that of the primary. {\it The
separation of the primary and secondary components, $1.5 \leq \Delta
\theta \leq 50$~mas}.  This is set by the VLBI resolution and
field--of--view limitations and determines the mass range to which the
search is sensitive. The upper limit on the separation was set by the
practicalities of map--making at the time the VLBI surveys were
made. {\it The ratio of the flux densities of the core of the primary
component and the secondary component is $\leq$ 40:1}. At this level
secondaries are detected with high signal--to--noise ratio.

Fifty lens candidates were selected of which six could immediately be
ruled out from other published VLBI data.

\section{Scrutiny of Lens Candidates}

Since the brightness contrast between the core and all other emission
regions in a radio source increases with increasing frequency, the
simplest way to determine whether two or more of the components are
compact, and have properties appropriate for lensed images, is to
observe at a higher frequency and higher resolution than the original
5 GHz survey maps. The candidates were therefore all observed at 15
GHz using the Very Long Baseline Array (VLBA).  A minority of the
candidates were also observed at one or more of the frequencies 1.6,
8.4 and 22 GHz. The observations and the radio maps will be described
elsewhere.

\begin{figure} 
\setlength{\unitlength}{1.0cm}
\begin{picture}(8.5,8.5)(0,0) 
\put(-1.3,-4.6){\includegraphics{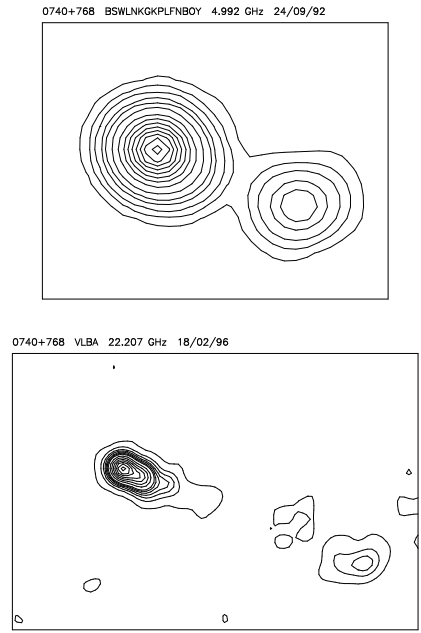}}
\end{picture} 
\caption{{\em Upper)} VLBI map of a millilens candidate 0740+768 at 5
GHz and 1 milliarcsec resolution; the separation between the two
components is 3.2 mas. {\em Lower)} VLBI map of 0740+768 at 22 GHz and
0.3 milliarcsec resolution; the two components, barely resolved at 5
GHz, have markedly different surface brightnesses. Rather than a
characteristic lensing morphology 0740+768 displays the asymmetric
``core--jet'' structure commonly found in flat--spectrum radio
sources.}
\label{fig1}
\end{figure}

Many of the candidates turned out to be core--jet sources in which the
jet is faint but contains a ``hot--spot''. The VLBA 15-- and 22--GHz maps
reveal the bright core and resolve the lower-brightness hot-spot.  An
example of such a candidate is shown in Figure \ref{fig1}. The 5 GHz
map of 0740+768 shows a simple unresolved double source but the higher
resolution 22--GHz map clearly shows that the compact core lies at the
eastern (left) end with the other ``jet-like'' emission having
markedly lower--brightness. Other candidates turned out to be Compact
Symmetric Objects (CSOs)---an early ($<10^{4}$ years) high--luminosity
phase of large double radio sources \cite{Rea96}. At high resolution
CSOs often show a single central core straddled by a pair of outer
``lobes'' of significantly lower surface brightness; not only is there
no detailed correspondence between sub--structures in the lobes but
also their parities are incorrect for lensed images.

Component spectra can also be used to rule out candidates. For lensing
on the present scales the time delay between the light paths ranges
from a few seconds to a few hours, which is much shorter than the
timescales between sets of VLBI observations.  Lensed images must,
therefore, brighten and fade together. A strong corollary is that
lensed images must have identical radio spectra or, equivalently,
their flux ratio must not be frequency--dependent. This argument
should be used with care if one or both of the components is well
resolved but we ruled out candidates with compact components if the
flux ratios at different frequencies were inconsistent by $\gta 30$\%.

Component motion is also an excellent discriminant. If the lens moves
across the line of sight the relative separation of the images will
change. For example a lens at $z \sim 0.5$, moving transversely with
$v \sim 1000$ km s$^{-1}$, will produce a relative image motion $\sim
1 \times 10^{-5}$ mas yr$^{-1}$. This is much too small to be
measurable with VLBI and thus, if relative motion is detected, the
secondary cannot be a lensed image. In a separate follow--up programme
for the PR and CJ VLBI surveys most of the sources have been observed
more than twice over a period of several years \cite{Bri00}. Currently
over 30 of our 50 candidates are known to show component motions
greater than $\sim 0.1$ mas yr$^{-1}$ and hence can be ruled out by
this means as well.

All 50 candidates were ruled out, often for a combination of the
reasons cited above, and we now use this null result to set
quantitative limits on supermassive \CO for objects {\it uniformly
distributed} throughout the universe.

\begin{figure} 
\setlength{\unitlength}{1.0cm}
\begin{picture}(8.5,8.5)(0,0) 
\put(-0.6,9.5){\includegraphics{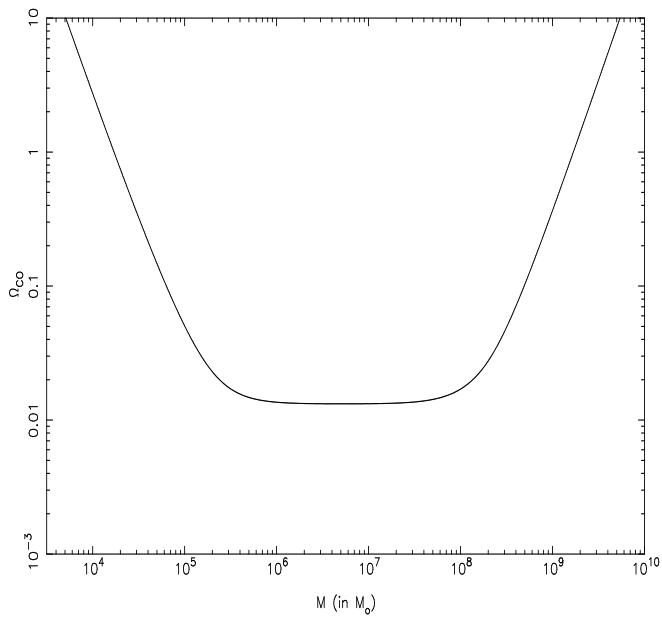}} 
\end{picture} 
\caption{Limit on \CO at the 95\% confidence level from the failure to
detect gravitational lensing amongst 300 compact sources with image~
separations in the range 1.5--50 milliarcsec.}
\label{fig2}
\end{figure}

\section{Limits on \CO}

We used the detection volume method introduced by Nemiroff
\cite{Nem89} and developed in detail for the specific case of a VLBI
search for point masses by KKB. We will, therefore, only make a few
points specific to our search. The most important observational limit
is the flux ratio $R(<40:1)$ which sets the maximum source--lens
impact parameter and hence defines the basic detection volume.  The
usual definition of ``strong'' lensing \cite{Tur84} refers to images
with a flux ratio of $<7:1$; our $<40:1$ flux ratio therefore
corresponds to a configuration with a larger lens--to--source ``impact
parameter'' and hence the cross-section for lensing is almost five
times that normally assumed for lensing calculations. Each source is
observed with an approximately fixed angular resolution and hence
minimum image separation $\delta(=1.5)$~mas, and limited
field--of--view $\Delta(=50)$~mas, both of which act to truncate the
detection volume. In effect $\delta$ and $\Delta$ define the lower and
upper limits of a mass range for which the truncation is small. Our
search is sensitive to the mass range $\sim 10^6$ to $\sim 10^8$
M$_{\odot}$. For observations of a number of sources the individual
detection volumes for each source can be summed to give a total
detection volume.  In the calculation of these volumes one needs to
know the source redshifts and these are available for 270 out of the
300 sources in our sample; the mean redshift is 1.30 and 20\% of the
sources have redshifts $\geq 2$. For the remaining 30 sources we have
assumed a distribution in redshift similar to that exhibited by the
other 270. For the calculation of the angular size distances we took
$H_0=65$ km s$^{-1}$ Mpc$^{-1}$ and, for direct comparison with KKB's
result we assumed an Einstein--de Sitter Universe ($\Omega_M=1;
\Omega_{\Lambda}=0$).

Our null result allows an upper limit to be placed on \CO and
Figure \ref{fig2} shows the limit at the 95\% confidence level as a function
of CO mass.  The limit is approximately constant from $\sim 10^6$ to
$\sim 10^8$ M$_{\odot}$ and in this mass range \CO $\lta 0.013$. If
\CO was equal to 0.013 the ``expected'' number of lenses in the total
detection volume associated with our 300 sources would be three and
the fact that none are detected allows us to reject this hypothesis
with 95\% confidence. If \CO was equal to 0.026 then $\sim 6$ lenses
would be expected and we can reject this higher mass density with
99.7\% confidence.

These limits are conservative because gravitational lensing increases
the observed flux density of a background source and hence lensed
sources are drawn from a fainter source population than the unlensed
sources; a flux--limited survey will contain more lenses than
expected. Our null result therefore corresponds formally to a stronger
limit on \CO but the ``magnification bias'' associated with flat
spectrum radio sources, especially for lens systems with high flux
ratios, is of order unity \cite{Kin96} and we have therefore ignored
this effect.

\section{Discussion} 

KKB derived a limit on \CO ($10^7$--$10^9$ M$_{\odot}$) $< 0.4$
(99.7\% confidence); this corresponds to \CO $<0.2$ (95\%
confidence). The lower mass limit is larger than ours since they took
$\delta \sim 4$ mas compared with our 1.5 mas. The upper mass limit
derived by KKB is high since it is based on $\Delta=500$~mas which is,
in our view, optimistic; we used $\Delta=50$~mas. Our null result
implies \CO $\lta 0.013$ (95\% confidence) in the mass range $\sim
10^{6}$ to $\sim 10^{8}$M$_{\odot}$ which is about 15 times more
stringent than KKB's. The improvement arises from a combination of
more sources (300 cf. 48), larger $R$ ($<40:1$ cf. $<20:1$) and hence
larger lensing cross--section and a higher mean redshift for the
parent sample (1.30 cf. 0.89). Our result allows the following
conclusions to be drawn.

Uniformly distributed CO in the mass range $\sim 10^6$ to $\sim 10^8$
M$_{\odot}$ do not make up more than $\sim 1\%$ of the closure density
$\Omega_{total}=1$ which is strongly supported by the latest
measurements of the angular spectrum of the Cosmic Microwave
Background \cite{Jaf00}. Similarly such CO do not make up more than
$\sim 3\%$ of the Dark Matter density $\Omega_{DM} \sim 0.3$ favoured
by current observations. The favoured value of the baryon density from
Big--Bang Nucleosynthesis is $\Omega_{b}h^{2}=0.019 \pm 0.002$
\cite{Bur99}. Taking a plausible value for $h$ (0.65) implies
$\Omega_{b} = 0.045 \pm 0.005$ thus our 95\% confidence limit implies
that uniformly--distributed Jeans--mass CO do not make up more than
about one third of $\Omega_{b}$.

A large population of uniformly--distributed $\sim 10^{6.5}$
M$_{\odot}$ black holes is required by the cosmogonic model
developed by Gnedin \& Ostriker \cite{Gne92} and Gnedin, Ostriker \&
Rees \cite{Gne95}. This model predicts that up to $5\%$ of high
redshift sources should be detectably lensed by such a population \ie
there should be up to 15 lenses in our sample. They are not observed
and hence the model can be ruled out.

Perhaps the next interesting limit is \CO $\lta 0.005$ which would
constrain the contribution of supermassive CO to be no more than the
baryonic contribution of presently observable stars and galaxies. To
reach this limit about 1000 sources, with the same redshift
distribution as the present sample, would have to be studied; this
would be a time--consuming, but relatively straightforward, task.

We thank Silke Britzen for providing results prior to publication.
The VLBA is the Very Long Baseline Array and is operated by the
National Radio Astronomy Observatory which is a facility of the
National Science Foundation operated under cooperative agreement by
Associated Universities, Inc.

{\em Note added in proof}: When this work was almost completed we
became aware of the millilensing results of Nemiroff and collaborators
(preceding PR Letter) using gamma ray bursts. Their work has produced
limits on the CO abundance which are similar to ours.

\end{document}